\documentclass[twocolumn]{revtex4}
\usepackage{hyperref}

\usepackage[utf8]{inputenc}
\usepackage[T1]{fontenc}

\usepackage{xcolor}
\usepackage[]{graphicx}
\usepackage{subfigure}

\usepackage{braket}

\usepackage{import}

\usepackage{amsmath, amsfonts, amssymb, mathrsfs}

\newcommand{\up}{\uparrow} 
\newcommand{\down}{\downarrow}

\newcommand{\Tr}[1]{\text{Tr} \left[ #1 \right]}

\newcommand{\nvec}{\mathbf{n}}

\newcommand{\Hop}{\hat{H}}
\newcommand{\Uop}{\hat{U}}
\newcommand{\Vop}{\hat{V}}
\newcommand{\Kop}{\hat{K}}
\newcommand{\Nop}{\hat{n}}
\newcommand{\Cop}{\hat{c}}

\newcommand{\refeq}[1]{\!(\ref{#1})}

\usepackage{natbib}
\usepackage{comment}

\begin{document}\sloppy

\title{Melting a Hubbard dimer: benchmarks of `ALDA' for quantum thermodynamics}

\author{Marcela Herrera}
\affiliation{Centro de Ciências Naturais e Humanas, Universidade Federal do ABC, 
Avenida dos Estados 5001, 09210-580, SantoAndré, São Paulo, Brazil
}

\author{Krissia Zawadzki}
\affiliation{Departamento de Física e Ciência Interdisciplinar, Instituto de Física de São Carlos,
  University of São Paulo, Caixa Postal 369, 13560-970 São Carlos, SP, Brazil} 

\author{Irene D'Amico} \affiliation{Department of Physics, University of York, York, YO10\,5DD, United Kingdom}
\affiliation{Departamento de Física e Ciência Interdisciplinar, Instituto de Física de São Carlos,
  University of São Paulo, Caixa Postal 369, 13560-970 São Carlos, SP, Brazil}

\begin{abstract}
	 The competition between evolution time, interaction strength, and temperature challenges our understanding of many-body quantum systems out-of-equilibrium. Here we consider a benchmark system, the Hubbard dimer, which allows us to explore all the relevant regimes and calculate exactly the related average quantum work. At difference with previous studies, we focus on the effect of increasing temperature, and show how this can turn competition between many-body interactions and driving field into synergy.  
We then turn to use recently proposed protocols inspired by density functional theory to explore if these effects could be reproduced by using simple approximations.
We find that, up to and including intermediate temperatures, a method which borrows from ground-state adiabatic local density approximation improves dramatically the estimate for the average quantum work, including, in the adiabatic regime, when correlations are strong. However at high temperature and at least when based on the pseudo-LDA, this method fails to capture the counterintuitive qualitative dependence of the quantum work with interaction strength, albeit getting the quantitative estimates relatively close to the exact results.
 \end{abstract}


\maketitle

\section{\label{sec:intro} Introduction}
	In the last decades, we have witnessed remarkable progress in Density
Functional Theory (DFT) with the development of various tools
to study many-body quantum systems \cite{Yang2012,Burke2012,Perdew2013,Jones-RMP.87.897,Yu2016}.
In particular, time dependent DFT (TD-DFT) \cite{Runge-PRL.52.997}
allowed to go beyond  ground-states and  opened
a new path \cite{Burke-JCP.123.6,Casida2012,Ullrich2011} in the ab initio formulation
for electronic transport \cite{Kurth-PRB.72.035308,Karlsson-PRL.106.116401,Brandbyge-PRB.65.165401}, electronic excitations \cite{Ullrich2011},
and calculations of thermodynamical properties of solid state systems
\cite{Smith-FQC.978-981-10-5651-2,Herrera-SR.07.4655}. Currently, with the urge to describe accurately realistic devices for quantum technologies, theoretical and computational physicists have been devoting great efforts to improve the treatment of out-of-equilibrium systems
\cite{Brandbyge-PRB.65.165401, Pribram-Jones-PRL.116.233001}.

As in the classical world, thermodynamics will impose limits in the
fabrication and operation of devices for quantum technologies~\cite{Goold2016,Vinjanampathy2016,Millen2016,Parrondo2015,Girolami2015}.
The descriptions and the laws as formulated within the conventional thermodynamics
are no longer valid at the scales where these technologies are being
developed~\cite{Castelvecchi2017}. At this level energy fluctuations
become important, and quantities such as work, heat, and entropy production
are treated as stochastic variables~\cite{Bustamante2005}. The quantum
thermodynamics research has been fueled recently by experiments,
on small and non-interacting systems ~\cite{Batalhao2014,An2014,Batalhao2015,Peterson2016,Camati2016,Micadei2017}. However, the experimental study of
the thermodynamics of quantum many-body systems remains a challenging task, and
even theoretical studies could require an enormous computational power.
The investigation of the thermodynamics of the emergent collective
phenomena in quantum many-body systems is, without a doubt, a fascinating subject. In this context some discussions about quantum
thermodynamic properties in out-of-equilibrium systems have been
reported (see e.g. \cite{Silva2008,Mascarenhas2014,Fusco2014,Zhong2015,Eisert2015,Leonard2015,Solano-Carrillo2016}).
In particular, a method to calculate quantum thermodynamic properties of interacting systems subject to driving fields  was presented
recently in Ref.~\cite{Herrera-SR.07.4655}, where it was applied to the calculation of the  average quantum work in a Hubbard
dimer driven by a time-dependent external potential. Such method uses an
approximated framework based on tools from DFT, where the many-body
problem dynamics is mapped onto the non-interacting dynamics of Kohn-Sham Hamiltonians.
As it is shown in ~\cite{Herrera-SR.07.4655}, DFT can offer a reliable
approach for estimating thermodynamical properties in out-of-equilibrium
systems: this method provides  good accuracy when
a suitable approximation for the so-called exchange-correlation
potential is used.

Using the approach in~\cite{Herrera-SR.07.4655},
the present paper aims to examine in more detail the use of different levels
of approximation for the proposed DFT-protocol.
In particular, we focus on the effect of increasing temperatures, and discuss the competition between temperature, coupling regime
and evolution time on the average quantum work $W$ extracted from a two-qubit
system subject to a non-linear dynamics. We will compare the exact $W$ extracted from
the interacting system with its estimate from the non-interacting
picture, and from approximations based on the local density approximation (LDA).
We will examine the improvement in the calculations
when the dynamical effects are included into the exchange-correlation
functional by means of an adiabatic LDA-inspired method (ALDA-i) and explore the limits of a ground-state formalism as the system temperature rises.
 
\section{\label{sec:dimer} The driven Hubbard dimer}
		
	The two-site Hubbard Hamiltonian is widely used for benchmarking density-functional approaches for strongly correlated systems \cite{Herrera-SR.07.4655,Carrascal-JPCM.27.393001,Fuks2014}. It also provides a simplified model to study two qubit systems in all coupling and dynamical regimes. The size of the problem allows for exact solution as well as, potentially, for experimental verifications. The Hubbard dimer has in fact been simulated using different types of systems, such as quantum dots \cite{Coe2010,Kikoin2012,Barthelemy2013} and cold atoms \cite{Murmann2015}, and could be simulated using small molecules driven by NMR techniques.

	An interacting Hubbard dimer driven by a non-homogeneous potential
	$\Vop(t)$ is described by the Hamiltonian
	\begin{equation}
		\label{eq:ham_Hubbard_dimer_inhom}
		\Hop (t) =  \Kop + \Uop + \Vop (t).
	\end{equation}
	
    Here
	\begin{align}
		\label{eq:H_hopping}
		\Kop = - J \sum_{\sigma} (\Cop_{1\sigma}^\dagger \Cop_{2\sigma} + \Cop_{2\sigma}^\dagger \Cop_{1\sigma} )
	\end{align}
represents the kinetic energy with hopping parameter $J$;
$c_{i\sigma}^\dagger$ ($c_{i\sigma}$) is the creation (annihilation) operator acting on sites $1$ and $2$ with spin $\sigma = \up, \down$;
	\begin{align}
		\label{eq:H_cou}
		\Uop =  U ( \, \Nop_{1\up} \Nop_{1\down} + \Nop_{2\up} \Nop_{2\down} )
	\end{align}
	represents the electrostatic Coulomb interaction of strength $U$, and number operator $n_{i\sigma} = c_{i\sigma}^\dagger c_{i\sigma} $, with $i=1,2$; and, finally,
	\begin{align}
		\label{eq:Vt}
		\Vop(t) = V_1(t) \Nop_1 + V_2(t) \Nop_2
	\end{align}
	is the external, non-homogeneous, time-dependent potential.

\section{\label{sec:dynamics} Non-linear Dynamics and Quantum Work}	
    
\begin{figure}
    	\begin{center}
        \includegraphics[width=0.95\columnwidth]{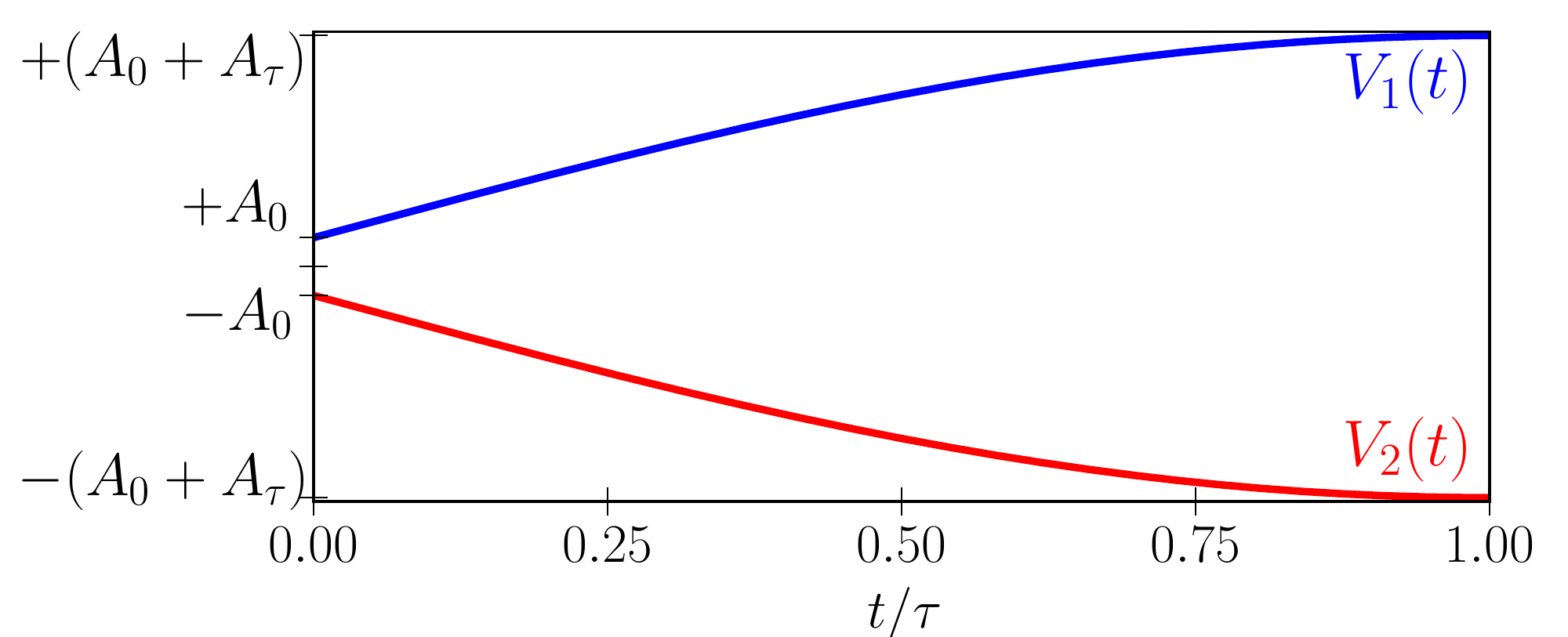}
         \end{center}
         \caption{\label{fig:Fig1} Non-linear sinusoidal potential from Eq. \refeq{eq:Vextsin}. During the dynamics, the driving potential promotes a flux of electrons from site $1$ to site $2$. When the system becomes interacting ($U \neq 0$), the site-occupations $n_1$ and $n_2$  at $t=\tau$ will depend not only on the parameters $A_0$ and $A_\tau$ of Eq. \refeq{eq:Vextsin}, but also on the coupling $U$, the temperature $k_B T$ and the hopping parameter $J$.}
	\end{figure}

    	In our analysis, we consider a two-site Hubbard model driven by a non-linear potential with sinusoidal form
	\begin{align}
		\label{eq:Vextsin}
		V_j(t) & = (-1)^j \left[ A_0 + A_\tau \sin\left( \omega_{4\tau} t \right) \right],
	\end{align}
	where $j=1,2$ labels each site. In this paper, we consider the parameters $A_0 = J$ and $A_\tau = 7J$ and explore coupling regimes from fully non-interacting ($U/J = 0$) to the strongly coupled ($U/J \approx 10$). The potential is plotted in Fig.~\ref{fig:Fig1}.

	We will focus on the extraction of quantum work $W$ during a dynamical process at finite temperature which starts at $t=0$ and ends at $t=\tau = \pi/(2\omega_{4\tau})$.  At $t=0$, the system is described by the equilibrium thermal state
    \begin{align}
    	\label{eq:thermal_state_0}
        \rho(t=0) & = \frac{e^{-H(t=0) / k_B T}}{Z(t=0)},
    \end{align}
    where $Z(t) = \Tr{e^{-H(t) / k_B T}}$ is the instantaneous partition function, $k_B$ is the Boltzman constant and $T$ is the temperature.
Apart from the driving field, the system is considered isolated. At the end of the process, $\rho(t=\tau)$ is not, in general, described by an equilibrium state.

    The extracted quantum work $W$ is defined in terms of the work probability distribution $P(w)$, a thermodynamical quantity yielding information about the energetic transitions and non-equilibrium fluctuations taking place while the quantum system evolves through a driven dynamics.
We define $W$ according to \cite{Talkner2007} as
    \begin{align}
        \label{eq:quantumWork}
         W  & = - \int w \, P(w) \, dw,
    \end{align}
    where $P(w)$ is given by
    \begin{align}
        \label{eq:quantumWorkdistribution}
        P(w) = \sum_{n,m} \, p_n(t=0)  \, p_{m(t=\tau)|n(t=0)} \, \delta(w-\Delta \epsilon_{m,n}).
    \end{align}

    Here $p_n(t=0)$ denotes the population of the $n$-th eigenstate $\ket{\psi_n(t=0)}$ of the initial Hamiltonian; $p_{m(t=\tau)|n(t=0)}$ defines the probability to find the final system $\rho(t=\tau)$ in the $m$-th eigenstate $\ket{\psi_m (t=\tau)}$ of the final Hamiltonian $\Hop(t=\tau)$, when starting in $\ket{\psi_n (t=0)}$; and $\Delta \epsilon_{n,m}$ is the energy difference between eigenenergies $E_n(t=0)$ and $E_m(t=\tau)$.

    Let us examine energy, temporal and thermal scales of the problem. The competition between $J$, $U$ and the strengths of the external potentials $A_0$ and $A_\tau$ dictates the first. In the weakly-coupled regime $U \ll J$, one might expect a non-interacting description to be adequate for representing the system. The changes in the instantaneous densities $n_j(t)$ are ultimately determined by a delicate balance with the external time-dependent potential. For increasing values of $U$, the external potentials, transforms the second site in an electron sink, and enters the competition with the Coulomb repulsion.

\begin{figure}[htb!]
    	\begin{center}
        \includegraphics[width=1.125\columnwidth]{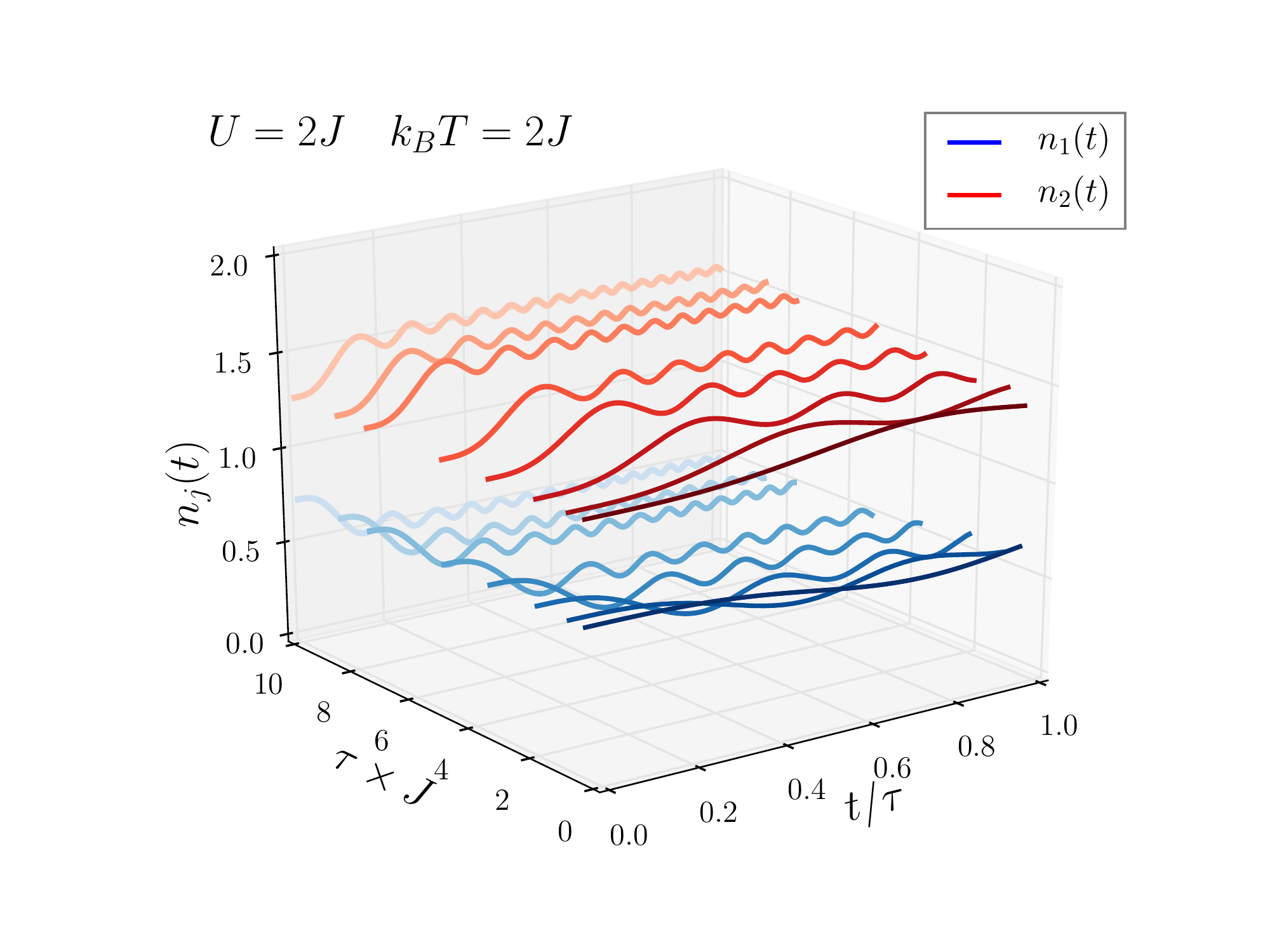}
         \end{center}
         \caption{\label{fig:Fig2} Evolution of site occupations $n_j(t)$ ($j=1,2$) versus the rescaled time $t/\tau$ for dynamics driven by the parameters $U=2J$, $A_0=J$, $A_\tau = 7J$ and different evolution times $\tau \times J= 0.5, ~1.0, ~2.0, ~3.5, ~5.0, ~7.5, ~8.5, ~10.0$. The darker the lines the closest the dynamics is to a sudden quench, whereas the lightest color depicts a regime close to adiabatic.}
	\end{figure}

  The parameter $\tau$ controls the speed at which the particles are driven. For $\tau \times J \ll 1$, the dynamics corresponds to a sudden quench: the evolution is so fast that the system does not have enough time to adapt. At the opposite limit, $\tau \times J \gg 1$, the evolution can be considered adiabatic. The dynamic regimes are illustrated in Fig. \ref{fig:Fig2}, where we show the instantaneous densities $n_1(t)$ (red tones) and $n_2(t)$ (blue tones) for $\tau \times J = 0.5$ (dark shades) to $\tau \times J = 10$ (light shades), $U = 2J$, and $k_B T = 2J$.
  In the sudden quench regime the system dynamics is so out of synchronicity  with the applied potential that the final value of the site occupations will strongly fluctuate with $\tau$.
  As the adiabatic regime sets in, the average value reached by the site occupation probabilities becomes much better defined, though a finer oscillation persists, due to the competing effects on electronic transport of external potential and Coulomb repulsion. The amplitude of this oscillation decays with increasing $\tau$.

\begin{figure}[htb!]
    	\begin{center}
        \includegraphics[width=0.925\columnwidth]{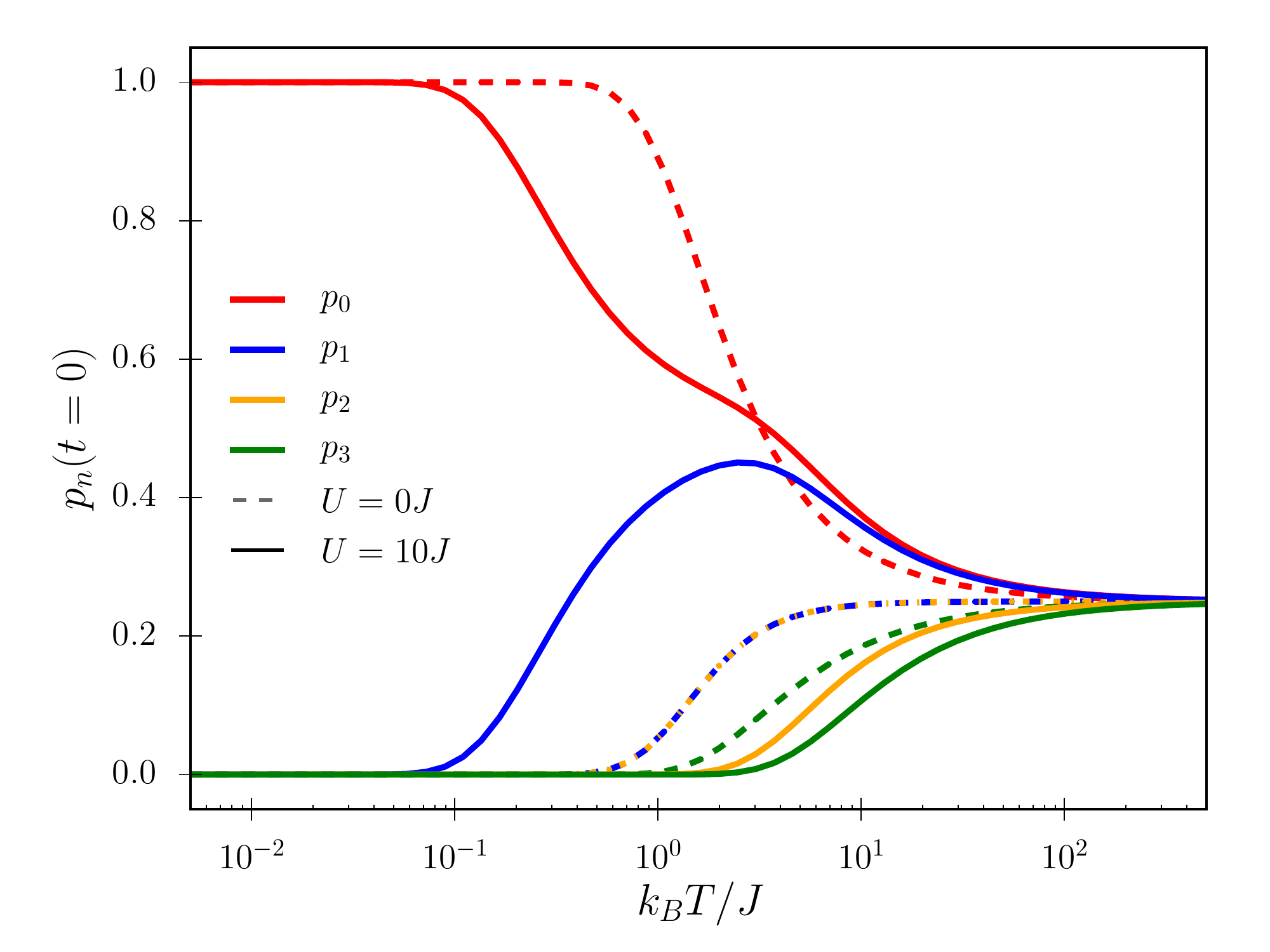}
         \end{center}
         \caption{\label{fig:Fig3} Thermal dependence of the populations $p_n$ ($n=0,1,2,3$) of each eigenstate $\ket{\Psi_n}$ of the initial Hamiltonian $\Hop(t=0)$ in the absence of coupling (dashed lines) and in the strongly coupling regime ($U = 10J$).}
	\end{figure}

The temperature introduces an additional dimension to the system parameter space.
Figure \ref{fig:Fig3} depicts the populations $p_n$ ($n=0,1,2,3$ for the dimer at half-filling) of the initial thermal state $\rho(t=0)$ as a function of the temperature $T$ (in units of $J / k_B$)
for $U/J = 0$ (non-interacting) and $U/J = 10$ (strongly-coupled). At very low temperatures, the initial thermal state corresponds to the pure ground-state $\rho(t = 0)_{T \rightarrow 0} = \ket{\psi_0} \bra{\psi_0}$.
Increasing temperature allows the system to populate different levels $E_n(t)$, giving rise to larger number of potential transitions during the dynamics. At very high-temperatures, all states equally contributing to the initial thermal state and a non-interacting picture arises.

\section{Zero-order DFT protocol}
	In this section we review the methods proposed in \cite{Herrera-SR.07.4655} to calculate the thermodynamic properties of an out-of-equilibrium system.

We write the interacting Hamiltonian as
\begin{align}
        \label{eq:interactingHam}
        \hat{H}(t) & = \hat{H}^{KS}(t) + \Delta \Hop (t),
    \end{align}
 with $ \Delta \Hop (t)\equiv  \hat{H}(t) -\hat{H}^{KS}(t)$ and  $\hat{H}^{KS}(t)$ the standard Kohn-Sham Hamiltonian~\cite{Jones-RMP.87.897}.

The `\emph{zero-order} DFT protocol' \cite{Herrera-SR.07.4655} approximates to zero-order the interacting Hamiltonian  with  $\hat{H}^{KS}(t)$. The advantage over the standard zero-order perturbation theory, is that $\hat{H}^{KS}(t)$ is formally non-interacting, but contains some of the interaction effects. This method has proven useful in estimating both entanglement \cite{Coe2008} and quantum thermodynamic quantities~\cite{Herrera-SR.07.4655}. In respect to the latter, an additional advantage of this method is that it formally avoids having many-body operators acting during the quantum evolution, which simplifies both the numerical simulations as well as potential experiments: these, in appropriate regimes, could be designed based on this approximation~\cite{Herrera-SR.07.4655}.

 Within the '\emph{zero-order} DFT protocol',   the initial thermal state is approximated
   as $\rho(t=0)  \approx \rho^{KS} (t = 0)$, with
   \begin{align}
   	\label{eq:rho0_KS}
    \rho^{KS} (t = 0) & = \frac{e^{-\Hop^{KS}(t=0) / k_B T}}{Z^{KS}(t=0)}.
   \end{align}

   Afterwards the system evolves according to the Kohn-Sham Hamiltonian $\Hop^{KS}(t)$, and at $t=\tau$ an approximation for the final state $\rho (t = \tau)$ can be extracted.  By means of this protocol, we can obtain an estimate for the average quantum work discussed in Section \ref{sec:dynamics}.

    For the driven dimer, the interacting system $\Hop(t)$ is mapped into a non-interacting system according to
	\begin{equation}
		\label{eq:ham_Hubbard_dimer_KS}
		\Hop^{KS} (t) =  \Kop + \sum_{j=1,2}\Vop_j^{KS} (t),
	\end{equation}	
	where the last term is the Kohn-Sham potential, defined as
	\begin{align}
		\label{eq:v_KS}
		\Vop_{j}^{KS} (t) & =  \left[V_{j}^{xc}[\nvec]( t; U) +  V_{j}^{H}[\nvec]( t; U) + V_j(t) \right]\Nop_j
	\end{align}		
    where $\nvec(t) = (n_1(t), n_2(t))$, and $ V_{j}^{H} = \frac{U}{2} n_j$ is the Hartree potential accounting for the classical electrostatic repulsion.

	The first terms on the right-hand side of eqs. \refeq{eq:v_KS} is an effective single-particle potential which accounts for exchange and correlation effects, the so-called exchange-correlation potential \cite{Kohn-PR.140.A1133}. In general, the exact functional form of the exchange-correlation potential is unknown, requiring approximations.
In this paper we will consider local density-type of approximations, as described below.

\subsection{\label{subsec:pLDA} Zero-order DFT protocol with Pseudo-LDA (p-LDA)}

We consider {\it time-independent} $ V_{j}^{xc}$ and $ V_{j}^{H}$.
They are calculated at time $t=0$, and $ V_{j}^{xc}$ through the pseudo-local density approximation \cite{Capelle2013}. They are given by
\begin{align}\label{eq:vxcH-pLDA}
V_j^{xc,p-LDA} & =  - 2^{-4/3}\frac{4}{3} U n_j^{\frac{1}{3}}\\
 V_{j}^{H,0} & = \frac{U}{2} n_j(t=0).
\end{align}

Although expression \refeq{eq:vxcH-pLDA} comply with the main idea of an ab-initio approach to approximate the exchange-correlation potential, it is known to be a rough approximation for Hubbard-like Hamiltonians. As discussed in Ref. \cite{Capelle2013}, LDA approximations based on the many-body solution of the homogeneous Hubbard Hamiltonian, such as the Bethe Ansatz Local Density Approximation (BALDA) \cite{Lima-PRL.90.146402}, offers a more appropriate approach. For the purposes of the present paper, considering p-LDA instead of BALDA suffices for a qualitative analysis of out-of-equilibrium systems: our aim here is to test the improvement in density-functional calculations with the inclusion of temporal dependences in the exchange-correlation functional. Therefore, our comparison of static and adiabatic p-LDA illustrates the situation in which one does not have at hand the best approximation for $V_{xc}$.

\subsection{\label{sect0order} Zero-order DFT protocol with Adiabatic Local Density Approximation}

So far we have considered zero-order Hamiltonians $\hat{H}_{0}$ where particle-particle interactions were included at most through time-independent functionals of the initial site-occupation. However a more accurate representation of the driven system evolution should be expected by including time-dependent functionals. We take inspiration from the ground-state adiabatic LDA (ALDA) \cite{Runge-PRL.52.997} and include a time-dependence by considering the same functional forms as for the static DFT but calculated at every time using the instantaneous thermal site-occupation. This time-dependence is local in time.

To implement this protocol numerically, we use a self-consistent cycle to obtain the time-dependent site-occupation $n_{j}(t)$ at all times, and from there the corresponding functionals $V_{H,j}[n_{j}(t)]$ and $V_{xc,j}[n_{j}(t)]$ at all times. We iterate the protocol by running the dynamics several time, until convergency for the site occupation at all times is reached.
In details: we use as starting point the exact density at the initial time, i.e., $n_{j}^{(0)}(t)=n_{j}^{(exact)}(0)$. From this density we obtain the KS potentials $V_j^{KS,(1)}(t)=V_j^{KS,(1)}[n_{j}^{(0)}(t)]$ and therefore the Kohn-Sham Hamiltonian $\hat{H}_{KS}^{(1)}(t)=\hat{H}_{KS}^{(1)}[n_{j}^{(0)}(t),t]$. We evolve the system using this Hamiltonian and we obtain the state of the system $\rho^{(1)}(t)$. From this state we can calculate the next iteration for the site-occupation $n_{j}^{(1)}(t)=\text{Tr}\left[\rho^{(1)}(t)\hat{n}_{j}\right]$. Using this, we restart the cycle. This cycle is repeated until the convergence criteria $\displaystyle \sum_{0<t<\tau}\lvert n_{j}^{(k-1)}(t)-n_{j}^{(k)}(t)\rvert/M=10^{-5}$ is satisfied, where the time $[0,\tau]$ is discretized in $M$ different values of $t$.
In the present paper we use the pseudo-LDA approximation as a base to implement this time-dependent approach.

\section{\label{sec:exact_results}Temperature effects on average quantum work, exact results}
	
In this section we analyze how the exact extracted quantum work $W_{exact}$ is modified by increasing temperature. For the Hubbard dimer and $0\le U/J \le 10$, the gap $\Delta E$ between ground and first excited states is in the range $0.39 J \le \Delta E \le 2.82J$, depending on the value of $U$. To consider all regimes of interest, we then consider the three temperatures: $T=0.2J/k_B$, for which excited states are barely populated; $T=2J/k_B$, for which $k_B T$ is comparable to $\Delta E$ and the excited states start to be populated; and  $T=20J/k_B$ for which the thermal energy is the largest energy in the system and all states become comparably populated (see Fig.~\ref{fig:Fig3}).

\begin{figure*}
\begin{center}
    {\includegraphics[width=0.90\linewidth]{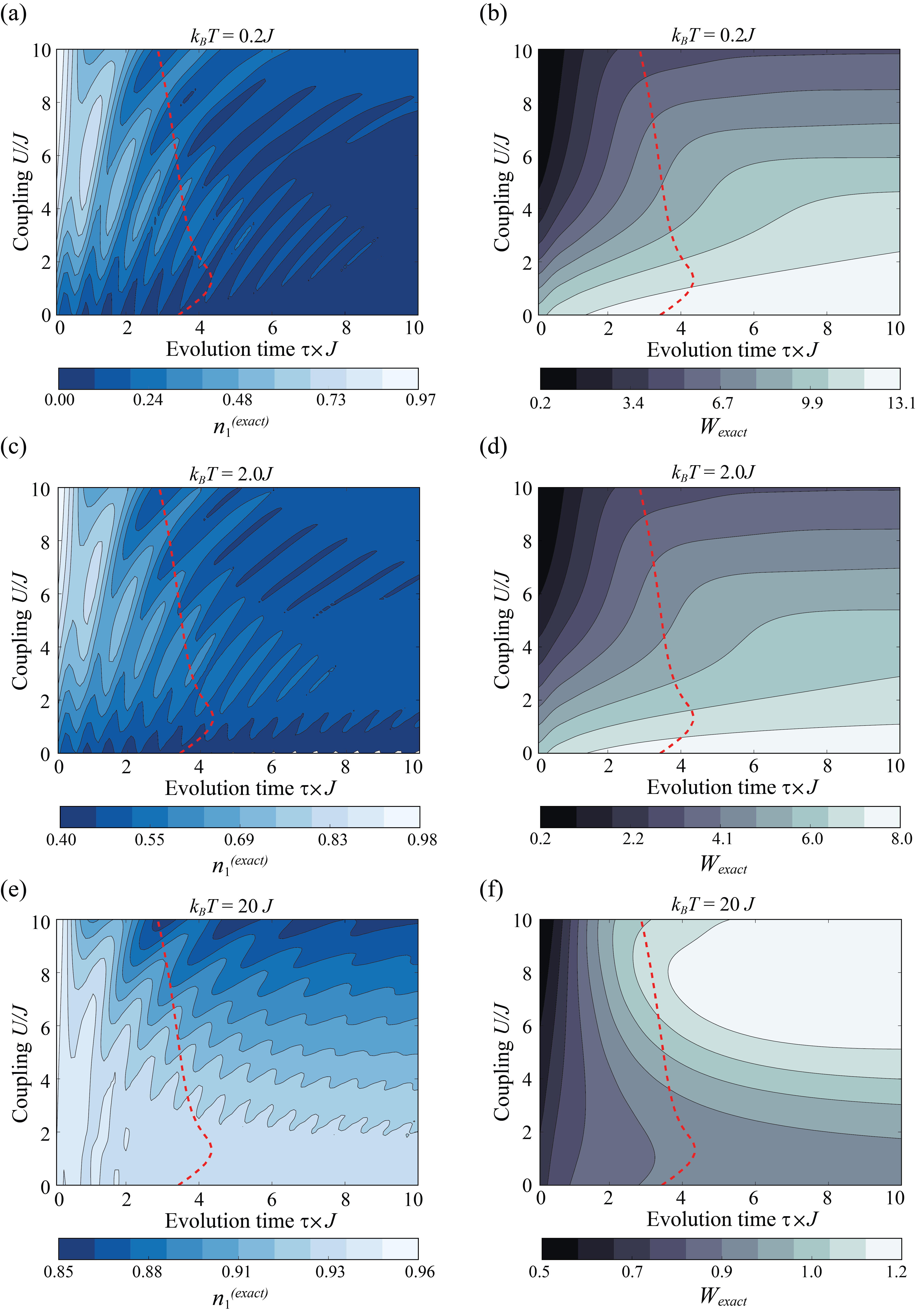}}
    \end{center}
         \caption{\label{fig:Fig4} Contour plots of the exact site-1 occupation at time $\tau$, $n_1^{exact}(\tau)$, (left) and the exact mean extracted quantum work $W_{exact}$ (right), with respect to the evolution time $\tau$ (x-axis) and the interaction strength $U$ (y-axis), for, top to bottom, $T=0.2J/k_B$, $T=2J/k_B$ and $T=20J/k_B$. The red dashed line indicates the transition region between sudden quench and adiabatic regimes.
         }
\end{figure*}

The results for $W_{exact}$ are shown in Fig.~\ref{fig:Fig4}, right column, where the contours lines for the quantum work are plotted against the Coulomb interaction strength $U$ and the evolution time $\tau$.  The hopping parameter $J$ is our unit of energy. It can be observed that for $k_B T \lessapprox \Delta E$ work can be mainly extracted in the low $U$ - large $\tau$ region (panels (b) and (d)), while when $k_B T >> \Delta E$ the situation is dramatically different, and the region with large $U$ becomes the most productive region (lower panel).

We explain this behavior as follows. In the sudden quench regime (small $\tau$ values, $\tau \times J < 1$), the system will not evolve significantly, $\rho(t=0)\approx\rho(t=\tau)$, and the quantum work will be in the great part determined by the characteristics of the eigenstates and spectra of the initial and final Hamiltonians, which are the same at all $T$. This explains why the variation of work production in this regime is mostly unaffected by temperature.
However the physics becomes very different as  $\tau$ increases, as the system becomes sensitive to the driving field, and eventually enters the adiabatic regime.

For $\tau \times J>1$ and $k_B T \lessapprox \Delta E$, the dominant contribution to the system thermal state comes from the ground state.
The dynamic of this will be affected by two competing forces: on the one side the driving field which drives the state towards double occupation in site 2; on the other the Coulomb repulsion which tends to decrease double occupation and, for very large $U$, would eventually lead to the precursor of the Mott-insulator phase transition.
For increasing $U$, the reduced efficacy of the applied field means that the field can extract increasingly less work from the system, as it is clearly observed in Fig.~\ref{fig:Fig4}, right column, panels (b) and (d). This is a noticeable effect of many-body interactions on the quantum work.
We note that the sign of the many body effect (reduction instead of increase of field efficacy) is due to the particular dynamics chosen, for which Coulomb repulsion and driving field are in competition. We would expect instead many-body interactions to help the process in the reverse dynamics: Coulomb interactions should help pumping work into the system when the system is driven from an initial state that favors double occupation towards a final state with a more homogeneous site potential. An example more subtle than the reverse dynamics will be given below.

This picture for $k_B T \lessapprox \Delta E$ is supported by the contour plots of the site occupation density $n_1(\tau)$ (panels (a) and (c) Fig.~\ref{fig:Fig4}). We see that for $\tau \times J \lessapprox 4$ and small $U$, $n_1(\tau)$ is depleted  by the action of the field well-below half-filling. As $U$ increases and the system is not yet adiabatic (compare to the red adiabatic line in the panels), the Coulomb repulsion strongly affects $n_1(\tau)$ by increasing its value. As $\tau$ increases further and the system settles into the adiabatic regime, the impact of the applied field increases, giving raise to the decaying ripples already observed in Fig.~\ref{fig:Fig2}.

Let us now focus on the bottom panels of Fig. ~\ref{fig:Fig4}. Here $k_B T >> \Delta E$ and $k_B T$ is also larger than the maximum difference in the site potentials induced by the applied field. Because of this, the effect of the dynamic regime becomes much less striking, as can be noted by the much reduced range of extracted work and site occupation variations (compare the scale of the contour lines between the bottom panels and the ones above them).  In addition, for the parameter considered, all states (ground state and excited ones) give comparable contributions to the thermal state. Also for all values of $U$ considered the inequality $k_B T \ge 2U$ holds, so that the effect of interactions becomes less important. Yet, quantitatively subtle, but qualitatively strikingly different behaviors from the ones observed in the upper panels, occurs and due to the presence of many-body interactions.

Counterintuitively and contrary to lower temperatures, the effect of increasing many body interactions is now to help the applied field to {\it deplete} site 1, as shown by the corresponding density contours.
In turn this enhances the field efficacy and hence the work that can be extracted  from the system. Maximum work can then be extracted in the adiabatic regime for large $U$ values (the red line within each figure indicates the transition region between the sudden quench and the adiabatic regime). In fact the adiabatic regime is the one in which the least entropy is produced, and hence the system is able to produce the largest work. This is confirmed by the results shown in Fig.~\ref{fig:Fig4}, panel (f).

The unexpected behavior of site occupation and average work with
increasing Coulomb interaction arises from the subtle interplay
between the evolution of the character of the eigenstates driven by
the applied field, and the substantial occupation of higher energy
states in the initial thermal state. When the evolution is adiabatic,
the final Hamiltonian eigenstates will inherit the occupation from the
initial thermal state. Strong Coulomb repulsion at $t=0$ implies that
eigenstates with strongly asymmetrical occupation are pushed further
up in energy so that, for the same temperature, at time $t=0$, higher
energy states become less populated for large $U$'s than for lower
values of $U$.

At strong coupling, the applied field must drive the system through an
anticrossing so that the final ground state at time $\tau$ may give
rise to the strongly asymmetric site occupation expected by a system
dominated by a step potential. So, in the adiabatic regime this
asymmetric state will have an increased weight at $t=\tau$ with
respect to what it had at $t=0$ due to the higher population of the
ground state at time $t=0$. The effect of this overall process is to
transfer population to the second site.

At lower $U$'s, for $U$ values less or comparable to the step in the
potential at $t=0$, the initial eigenstates are closer in energy as
they are not too influenced by the Coulomb repulsion. Because of this,
for the same temperature, the occupation of the highest energy state
will be larger than in the presence of strong coupling. After the time
evolution, this state corresponds to a distinctively asymmetric site
occupation, which favors site one. As the state maintains the high
thermal population acquired at time $t=0$, the effect of the overall
process is to transfer population to the first site. Note that, as the
highest excited state is involved, this process would be negligible at
low and intermediate temperature.

The combination of this high-temperature opposite population transfer
for weak and strong coupling leads to the behavior observed, where, at
high temperatures, increasing Coulomb repulsion favors asymmetry in
the $t=\tau$ site occupation.

In turn this enhances the field efficacy and hence the work that can be extracted  from the system. Maximum work can then be extracted in the adiabatic regime for large $U$ values (the red line within each figure indicates the transition region between the sudden quench and the adiabatic regime). In fact the adiabatic regime is the one in which the least entropy is produced, and hence the system is able to produce the largest work. This is confirmed by the results shown in panel (f) of Fig.~\ref{fig:Fig4}.

In the rest of the paper we will focus on the $T=2J/k_B$ and $T=20J/k_B$ cases, and study how various approximations capture the corresponding qualitative and quantitative changes in the production of quantum work.

\section{\label{sec:DFT-protocol}Accuracy of zero-order DFT-inspired protocols}

	\subsection{\label{subsec:intermediate-T}Intermediate-temperature regime, $k_BT=2J$}

\begin{figure*}
	\begin{center}
		\includegraphics[width=1.30\columnwidth]{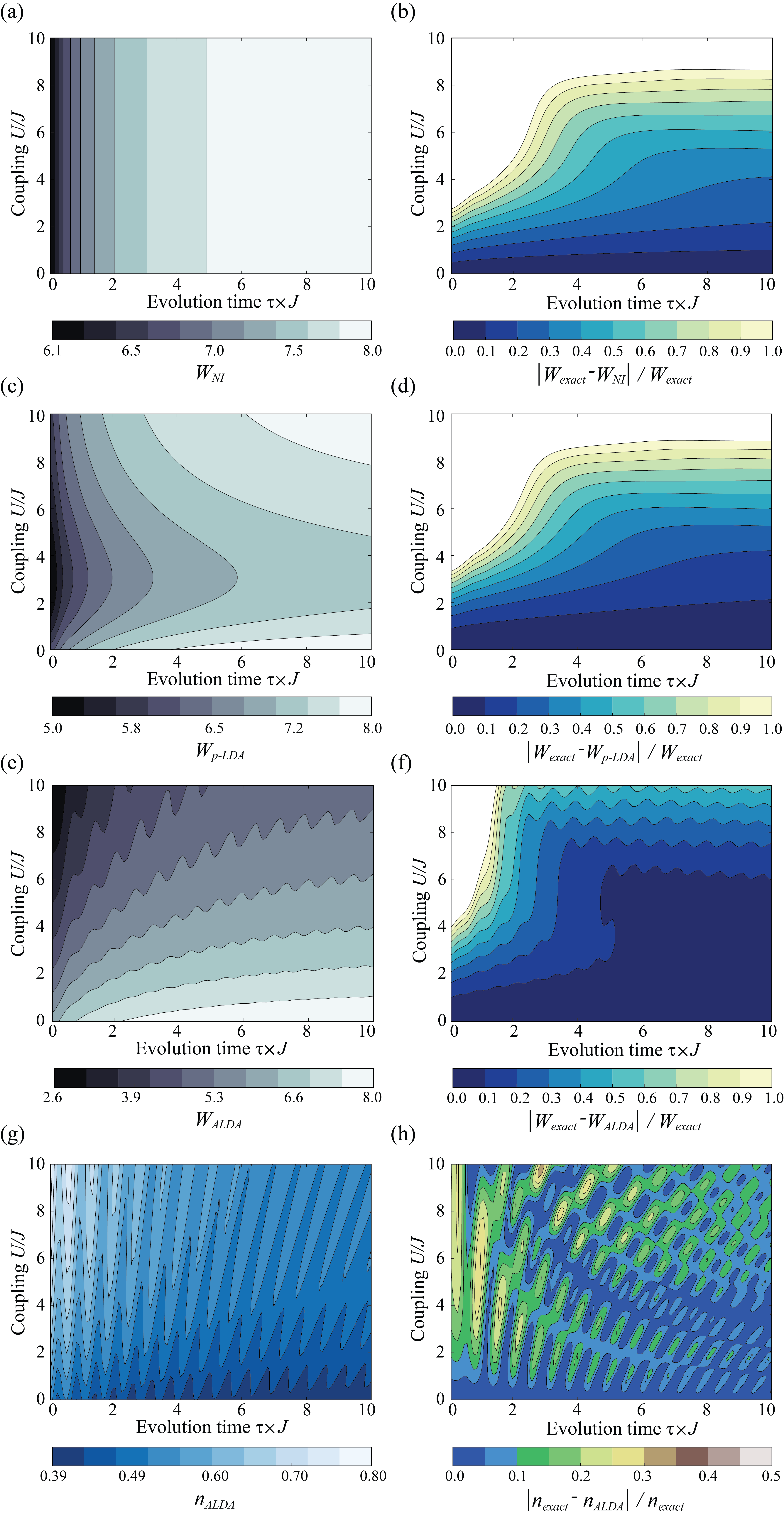}
	\end{center}
	\caption{\label{fig:Fig5} Upper three panels: Contour plots for the mean extracted quantum work $W$ (left) and its relative deviation from the corresponding exact values (right) for, top to penultimate panel, non-interacting, zero-order pseudo-LDA  and zero-order ALDA-inspired approximations at the intermediate temperature $T=2J/k_B$. Bottom panels:  Contour plot for zero-order ALDA-inspired site-one occupation at time $\tau$ (left) and its relative deviation from the corresponding exact values (right). All quantities are plotted against the evolution time $\tau$ (x-axis) and the Coulomb coupling $U$ (y-axis). }
\end{figure*}

In Fig.~\ref{fig:Fig5} we consider the effect of approximations when $k_BT=2J$.
In the upper three panels we show how climbing the ladder of `zero-order' protocols -- from the standard non-interacting, and through the DFT-inspired approximations -- improves the estimate of the average quantum work. From the top to the penultimate panel, we plot the approximated extracted work (left column) and its relative error with respect to $W_{exact}$ (right column) calculated from: a completely non-interacting Hubbard dimer ($U=0$); an interacting Hubbard dimer approximated by a Kohn-Sham Hamiltonian within the pseudo-LDA approximation; and finally an interacting Hubbard dimer approximated by a Kohn-Sham Hamiltonian within an ALDA-inspired approximation and based on the pseudo-LDA (see section \ref{sect0order}).

Let us first concentrate on the non-interacting results (upper panels of Fig.~\ref{fig:Fig5}). As should be expected, in this approximation the work is independent from $U$, and simply increases with $\tau$ as the system becomes more adiabatic. When compared to panel (d) of Fig.~\ref{fig:Fig4},  the non-interacting quantum work is qualitatively dramatically different from the corresponding exact one, with the exception of very small $U$ and especially in the adiabatic regime (Fig.~\ref{fig:Fig5} panel b).

When turning to the zero-order pseudo-LDA results ($W_{lda}$, Fig.~\ref{fig:Fig5} panel (c)), we observe that now the exact work is reproduced up to $U\approx 2 J$ both qualitatively and quantitatively. However, at larger $U$ values, as correlations become stronger, the match fails even qualitatively, compare Fig.~\ref{fig:Fig5} panel (c) with Fig.~\ref{fig:Fig4} panel (d).

While only a relatively minor improvement is noted between non-interacting and zero-order pseudo-LDA results, a remarkable improvement is observed when including time-dependent correlations through the ALDA-inspired scheme. Notably the bulk of this improvement is in the adiabatic parameter region, where non-Markovian effects are not important, in line with the fact that the ALDA approximation does not include memory effects. We stress though that in this intermediate-temperature regime the system state is clearly a mixed thermal state, while the scheme is based on an ALDA designed for pure states at zero temperature, so our result are non-trivial as in principle ALDA could badly fail for thermal states.

When considering $W_{alda}$ (Fig.~\ref{fig:Fig5} panel (e)), we notice that it qualitative reproduces the main features of the behavior of the exact work, and at all $U$ values. In addition the range of variation of $W_{alda}$ quantitatively closely match the one of $W_{exact}$, at difference with $W_{lda}$ which strongly overestimated the minimum amount of work (compare scales at the bottom of the related panels). As this is reached at large $U$'s, we consider this as a confirmation that this zero-order DFT-inspired approximation fairly treats correlation up to intermediate temperatures even when strong, at least for the system at hand. This approximation reproduces the exact work even quantitatively in the whole region corresponding to high extractable work ($W_{exact} > 5$), also a very good result.
We observe that the contour lines of panels (e) and (f) in Fig.~\ref{fig:Fig5} present oscillations and we will come back to this later in the paper.

In the bottom panels of Fig.~\ref{fig:Fig5} we consider the ALDA-inspired estimate for the site-one particle occupation at time $\tau$ (left) and show its relative difference with the corresponding exact result. We note that the site occupation has, on average, the best (worst) agreement in the same parameter region where the average quantum work is closely (farthest) reproduced.

\subsection{\label{subsec:high-T} High-temperature regime}

\begin{figure*}
	\begin{center}
		\includegraphics[width=1.30\columnwidth]{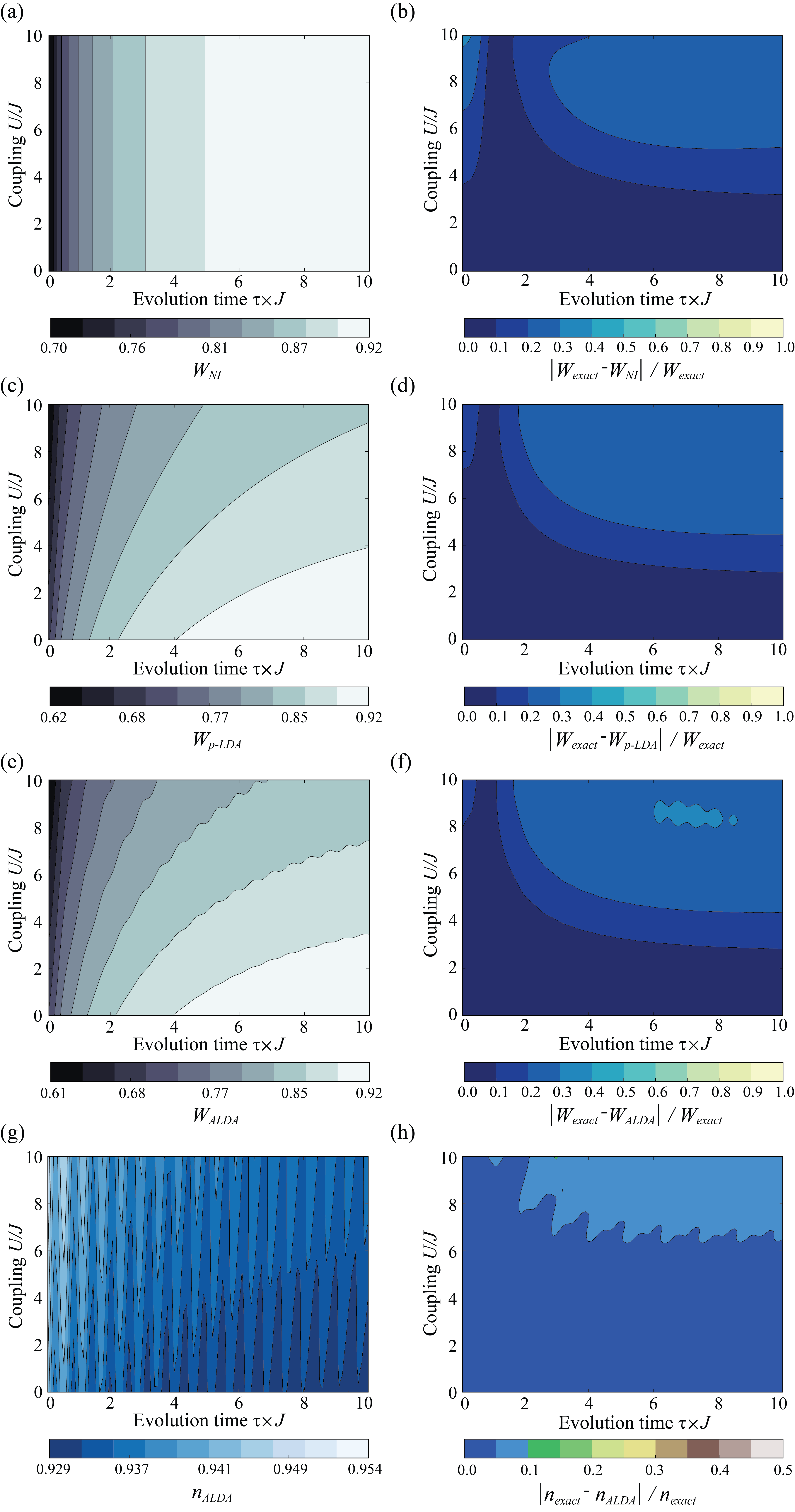}
	\end{center}
	\caption{\label{fig:Fig6} Upper three panels: Contour plots for the mean extracted quantum work $W$ (left) and its relative deviation from the corresponding exact values (right) for, top to penultimate panel, non-interacting, zero-order pseudo-LDA  and zero-order ALDA-inspired approximations at the high temperature $T=20J/k_B$. Bottom panels:  Contour plot for zero-order ALDA-inspired site-one occupation at time $\tau$ (left) and its relative deviation from the corresponding exact values (right). All quantities are plotted against the evolution time $\tau$ (x-axis) and the Coulomb coupling $U$ (y-axis). }
\end{figure*}

Fig.~\ref{fig:Fig6} shows the same quantities as Fig.~\ref{fig:Fig5} but for high temperature ($T=20J/k_B$).
By comparing the exact work (Fig.~\ref{fig:Fig4}, panel (f)) with the approximate estimates, we immediately notice that
all the approximations qualitatively reproduce the low $\tau$ regime ($\tau \times J  \lessapprox 2$) at all $U$ values (with the DFT-inspired approximations improving slightly over the non interacting one). However at the same time they all fail to reproduce the behavior, even qualitatively, at larger $\tau$'s and $U / J \gtrapprox 1$, as the regime becomes adiabatic and the interactions increase beyond perturbative. Interestingly enough, in this parameter region the corrections to the contour lines provided by the correlations introduced through the zero-order DFT-inspired approaches  go in the opposite direction with respect to the exact behavior.  The extent of extracted work is though comparable to the exact one, although its range is slightly smaller.

The above picture is well summarized in the second column of Fig.~\ref{fig:Fig6}, where the percentage variation with respect to the exact work is plotted, and the region where the exact work is worst reproduced indeed corresponds to the adiabatic, highly interacting parameter region.
We note that, because the value range of approximated and exact work are similar, the error never goes below 30\%.

Small differences between the results from the three approximations may be accounted for by noticing that the Hamiltonians remain different in the three cases, the more different the higher the value of $U$. We expect then small differences to appear for large $U$ values, as indeed is the case.

Panels (g) and (h) of Fig.~\ref{fig:Fig6} show the ALDA-inspired estimate for the site-1 occupation $n_1 (\tau)$ and its relative error with respect to the exact values, respectively. Panel (g) demonstrates that in this case, apart from the sudden-quench regime, the estimates from the ALDA-inspired zero-order approach behave {\it qualitatively} opposite to the exact behavior  (Fig.~\ref{fig:Fig4}, panel (e)) when parameters are varied.

We attribute this opposite behavior to the fact that in respect to the exact case, the instantaneous eigenvalues for the ALDA-inspired case are not as sensitive nor qualitatively nor quantitatively to the value of $U$. The gaps between higher excited states are poorly reproduced: any formally non-interacting Hamiltonian leads, for the Hubbard dimer, to a degeneracy between first and second excited states, while the widening of the gap between second and third excited states with increasing $U$ is underestimated by the ALDA-inspired estimates. This implies that the delicate interplay between thermal occupation and interaction strength at $t=0$, at the origin of the exact system behavior, is not reproduced.

Interestingly, due to the small variation of the site-one density values, {\it quantitatively} the overall picture is much better, as the relative error (Fig.~\ref{fig:Fig6} panel (h)) is actually very good in most parameter areas. Comparing this with the corresponding relative error for the average quantum work (panel (f)) confirms that there is a very good correlation between the quantitative accuracy of the ALDA-inspired estimates for local thermal density and average quantum work.

\subsection{\label{subsec:occTD} Site-occupation time-dependence for the ALDA-inspired scheme}

\begin{figure*}
	\begin{center}
		\includegraphics[width=0.90\linewidth]{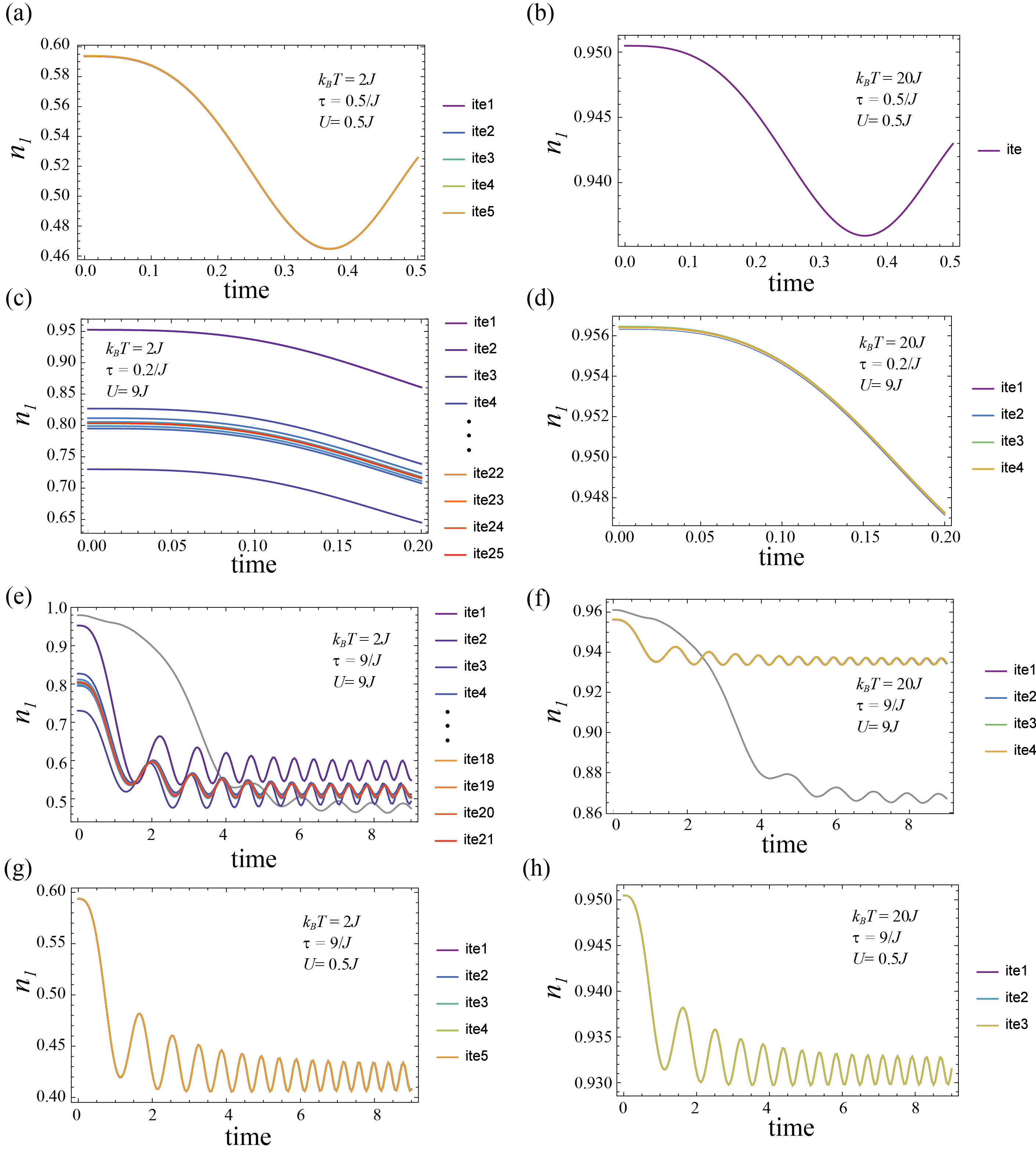}
	\end{center}

	\caption{\label{fig:Fig7} Zero-order ALDA-inspired site-one occupation estimate with respect to time for $T=2J/k_B$ (left) and $T=20J/k_B$ (right). Results from all the iterations necessary to achieve the accuracy of $10^{-5}$ are plotted as well. From top to bottom, calculations are done for the following points in the $\{\tau, U\}$ parameter space: $\{\tau,U\}=(0.5/J,0.5J)$; $\{\tau,U\}=(0.2/J,9J)$; $\{\tau,U\}=(9/J,9J)$; and $\{\tau,U\}=(9/J,0.5J)$. Number of iterations are indicated at the r.h.s. of each panel. The site occupation for the initial iteration was time-independent and chosen to be the one corresponding to the exact initial thermal state.
         }
\end{figure*}

In Fig.~\ref{fig:Fig7}  we show the time-dependence of the site-1  occupation probability $n_1(t)$ for the ALDA-inspired scheme. The left column refers to the intermediate temperature $k_BT=2J$ and the right column to the high temperature  $k_BT=20J$.
From top to bottom, we plot $n_1(t)$ for four points in the parameter space: $\{\tau,U\}=(0.5/J,0.5J)$ corresponding to  sudden quench and weak interactions;  $\{\tau,U\}=(0.2/J,9J)$ i.e. sudden quench and strong interactions; $\{\tau,U\}=(9/J,9J)$ adiabatic regime and strong interactions, and finally $\{\tau,U\}=(9/J,0.5J)$ corresponding to the adiabatic regime with weak interactions.

We note that, independently from $U$ and $T$, in the adiabatic regime, after an initial transient, the density oscillates around a mean value.
This explains the oscillations observed in panel (e) of Figs ~\ref{fig:Fig5} and ~\ref{fig:Fig6}.
These oscillations qualitatively reproduce the ones observed in the exact density, albeit with a different period (e.g. compare to the grey curve in Fig.~\ref{fig:Fig7} panels (e) and (f). This explains the oscillatory patterns observed in the panels (f) and (h) of Figs.~\ref{fig:Fig5} and \ref{fig:Fig6}. As in the exact case, see section 3, the oscillations are due to the interplay between the Coulomb repulsion and the attractive potential generated by the driving field, that induce transport in opposite directions.

Numerically, strong interactions (and the consequent `stiffness' of the system) slows down the convergency for the site occupation in the self-consistent cycle necessary to obtain the time-dependant KS potentials. This is most accentuated at low-intermediate temperatures as can be observed by looking at panels (c) and (e) of Fig.~\ref{fig:Fig7}, where a much larger number of iterations is necessary  to obtain the same level of convergency.

\section{\label{sec:conclusions} Conclusions}

	We have presented a study for the temperature dependence of the average quantum work that can be extracted from a driven Hubbard dimer at half-filling, for varying evolution time and Coulomb interaction strength parameters. The driving field was set up as to oppose, as the system evolved, the effect of inter-particle Coulomb repulsion,  by building a potential step between the two lattice sites at least twice as large as the maximum Coulomb interaction considered in this work. This competing behavior has allowed to uncover subtle and counterintuitive interplay between temperature, external driving field, and many-body effects on the production of quantum work.
On the one side, at low and intermediate temperatures and all other parameters the same, increasing Coulomb interaction strength indeed opposes the effect of the applied field, and, by doing so, reduces its efficacy and hence the work produced.
On the other side we discovered that, {\it at high temperatures} and medium and large system evolution times, Coulomb interaction strength {\it favors} the action of the applied driving field, and, consequently increases the production of work. We explain this counterintuitive behavior by the subtle interplay between interaction-dependent changes in the instantaneous eigenstates of the time-dependent Hamiltonian combined with the variation of their thermal occupation with temperature.
We also explore the possibility of approximating the exact work and the thermal site occupation by using zero-order DFT-inspired approximations. We find that {\it quantitatively} the ALDA-inspired approximation, based on the pseudo-LDA exchange-correlation potential, behaves well in most of the parameter space and at all temperatures considered.
However {\it qualitatively} the picture is more complex, and in particular even the ALDA-inspired approach, at least when based on the pseudo-LDA, fails to reproduce the subtle interplay between Coulomb interaction and temperature which, in the exact case, leads to the {\it depletion} of site-one for {\it increasing} Coulomb interaction at high temperature.

\section*{Acknowledgements}
MH acknowledges financial support from FAPESP (project no.2014/02778-1). KZ thanks CNPq (grant no. 140703/2014-4) and CAPES (grant no. 88881.135185/2016-01) for financial support. IDA acknowledges the Royal Society through the Newton Advanced Fellowship scheme (grant no. NA140436) and CNPq through the PVE scheme (grant no. 401414/2014-0).

\bibliographystyle{unsrt}
\bibliography{ref.bib}

\end{document}